\documentclass[aps,preprint]{revtex4-1}                                   
\usepackage{epsfig}

\begin{document} 
   
\title{Comparing different  protocols 
of temperature selection in the parallel tempering method}  
  
\author{Carlos E. Fiore} 
\email{fiore@fisica.ufpr.br} 
\affiliation{Departamento de F\'{\i}sica\\   
Universidade Federal do Paran\'a\\   
Caixa Postal 19044\\   
81531 Curitiba, Paran\'a, Brazil}   
\date{\today}   
   
\begin{abstract}  
Parallel tempering (PT) Monte Carlo simulations have been applied
to a variety of systems  presenting rugged free-energy landscapes.
Despite this, its efficiency  depends strongly on  
 the  temperature set. With this query in mind,
we present a comparative study among different 
temperature selection schemes in three  lattice-gas models. 
We  focus our attention in the constant entropy method (CEM), 
proposed by Sabo {\it et al}. In the CEM, the temperature  is chosen by
the fixed difference of entropy between adjacent replicas.
We consider a  method to determine the entropy which 
avoids  numerical integrations of the specific heat and other thermodynamic
quantities. Different analyses for first- and
second-order phase transitions have  been undertaken, revealing that
the CEM may be an useful criterion
for selecting the temperatures in the parallel tempering.
\end{abstract}  
 
 
  
\maketitle   
  
\section{Introduction}
Enhanced sampling tempering approaches, 
such as the parallel tempering (PT) \cite{nemoto}
and simulated tempering (ST) \cite{parisi},
 have played an important role in the study of complex systems such
as spin glasses \cite{spinglass}, protein folding \cite{proteins},
biomolecules  and others. The basic idea consists of using
the ``information'' obtained at
 high temperatures to systems at low temperatures, 
allowing the system to escape from metastable states and 
providing an appropriate visit of the configuration space.

Despite this, its efficiency  depends strongly on the temperature set.
In the last years, different procedures have been proposed  for the choice
of both the  temperature set and the number of replicas. 
Kofke \cite{kofke} has related the average acceptance probability
of replicas $\bar{p}_{\rm acc}$ with its difference of entropy.
Predescu {\it et al.} \cite{predescu}, based on the link
between acceptance rate and specific heat, argued that the optimal
distribution of intermediate temperatures should follow 
a geometric progression if the specific heat is nearly constant.
Kone {\it et al}. \cite{kone} proposed that  temperatures
in the parallel tempering  should be   
chosen  such a way that about $20\%$  of exchanges are accepted, 
when the  specific heat is also constant.
Katzgraber {\it et al}. \cite{helmut} considered a feedback 
optimized parallel tempering Monte Carlo that gives the temperature set
by measuring the replica diffusion through the temperature space. 
More recently Sabo {\it et al}. \cite{sabo} introduced the constant
entropy method (CEM), where
the temperature of replicas is chosen provided
the difference of entropy is held fixed.
An advantage of both CEM and feedback optimized methods is that
near the criticality temperatures become more concentrated, 
what increases the frequency of the replica exchanges when compared
with other schemes. Results for the Ising model \cite{helmut,sabo} 
revealed that they are equally efficient and superior than other schemes.
However, the CEM seems to be computationally 
simpler than the  method by Katzgraber {\it et al}.

In this paper, we give a further step
in order to ascertain the role  of the temperature schedule in the 
parallel tempering. We consider a comparative study 
among five different schemes for selecting the temperatures to be used:
arithmetic and geometric distribution of temperatures, arithmetic
distribution in inverse temperature \cite{calvo}, $ad-hoc$ distribution
of temperatures and the CEM.
In the {\it ad hoc} ensemble, temperatures are chosen in such a way that
 exchanges between adjacent replicas are about $30\%$ \cite{sabo,sabo2}.
In order to select the temperatures according 
to the CEM, we use the transfer matrix method \cite{sauerwein}, 
 in which the entropy
is obtained directly from Monte Carlo simulations, hence avoiding numerical 
integrations of thermodynamic quantities, such
as the specific heat and  energy. We  
shall  apply the comparative study to
different lattice models undergoing phase transitions
such as  Ising \cite{ising}, Blume-Emery-Griffiths (BEG) \cite{BEGMODEL}
and Bell-Lavis (BL) water  model \cite{bell}. 
Different analyses have been undertaken and 
for all schemes considered here the CEM
revealed to be  more advantageous.

This paper is organized as follows: In Sec. II we 
describe the  approaches employed here, in Sec. 
III we present the models, in Sec. IV we discuss
the numerical results and in Sec. V we give the conclusions.

\section{Parallel Tempering and the CEM }
The basic idea of the PT is that configurations 
from  high temperatures are used
to perform an ergodic walk in low temperatures. To 
this end, one
simulates simultaneously  a set of  $R$  replicas  ranged in the 
 interval  $\{T_1,...,T_{R}\}$ 
 by means of a Metropolis like algorithm \cite{metr} (although
in general the  interest lies in the lowest temperature  $T=T_1$). 
The actual MC simulation is composed of two parts: In the first part,
 a given site  $k$ of the  replica $i$
is chosen randomly and its  variable $\sigma_k$ may change to a new value
$\sigma_{k}'$ according to the Metropolis prescription
$p_i = \rm min\{1, \, \exp[-\beta_{i} \Delta \cal H]\}$, where
$\Delta{\cal H} = {\cal H}(\sigma') - {\cal H}(\sigma)$ is the
energy change due to the transition and 
$\beta_i = (k_{B} T_{i})^{-1}$.
In the second part, arbitrary pairs of replicas (say, at $T_{i}$
and $T_{j}$  with microscopy configurations $\sigma'$ and
$\sigma''$) can undergo  temperature switchings with probability
\begin{equation}                                                               
p_{i \leftrightarrow j} =                                                   
\min \{1, \, \exp[(\beta_{i} - \beta_{j})                                   
({\cal H}(\sigma') - {\cal H}(\sigma''))] \}.           
\label{p-pt}                                                                   
\end{equation}
In principle,  the replicas $i$ and $j$ may be adjacent
or not. Usually one adopts only adjacent 
exchanges, since the probability
of a given swap decreases by raising  $T$. In some cases, however,
 non-adjacent exchanges have revealed 
essential for the system  to escape from metastable states
\cite{juan,juan1,fiore8,fiore10,fiore11,calvo}.

As mentioned in the Introduction, we are going
to consider different procedures for determining the temperature
set. 
The former schemes (both arithmetic and geometric
schedules) are rather simple, since they 
require only the extreme temperatures  for determining
the whole set. 
In the {\it ad hoc} distribution, one
starts from $T_1$ for a given  $R$ and
 determines the $R-1$ temperatures in such a way that the exchange
probability between adjacent replicas is about $30\%$.
The CEM \cite{sabo} consists of adding intermediate
temperatures with  fixed difference of entropy.
More specifically,  given the
extreme temperatures $T_1$ and $T_{R}$ with entropies per volume
$s_1$ and $s_{R}$, respectively, we add
 $R-2$ intermediate temperatures 
$T_i$ whose  entropy $s_i$  is
 $s_i=s_1+(i-1)\times \Delta s$, where
$\Delta s=(s_{R}-s_1)/R-1$. 
Each value of  $s$  will be calculated
through  the thermodynamic equation $s=\frac{u-f}{T}$, where 
$u=\langle {\cal H}\rangle$ and $f$ is given by
\begin{equation}                                         
f=-\frac{1}{\beta V} \ln Z.  
\label{e28}                                                                   
\end{equation}
The transfer matrix method, used for 
 obtaining $f$ (and consequently $s$), is implemented by dividing 
a lattice with $V$ sites in $N$ layers  with $L$ ``spins''
($V=L\times N$). The  associated Hamiltonian is given by
\begin{equation}                                                               
{\cal H}= \sum_{k=1}^N {\cal H}(S_k,S_{k+1}),                                  
\label{e14}                                                                    
\end{equation}
where $S_k \equiv (\sigma_{1,k},\sigma_{2,k},...,\sigma_{L,k})$,
and due to the periodic boundary conditions $S_{N+1} = S_1$.

The probability $P(S_{1},S_{2},...,S_{N})$ of a given configuration
 is given by
\begin{equation}                                                               
P(S_{1},S_{2},...,S_{N})=\frac{1}{Z}                                           
T(S_{1},S_{2}) T(S_{2},S_{3})...T(S_{N},S_{1}),                                
\label{e15}                                                                    
\end{equation}
where  $T(S_{k},S_{k+1}) \equiv \exp ( - \beta {\cal H}(S_k,S_{k+1}))$
is an element of the transfer matrix $T$ and $Z=\rm Tr(\it {T^{N}})$
is the partition function.
The marginal probability distributions $P(S_{1})$ and $P(S_{1},S_{2})$
are given by
\begin{equation}                                                               
P(S_{1})=\frac{1}{Z} T^{N}(S_{1},S_{1}),                                        
\label{e17}                                                                    
\end{equation}
and
\begin{equation}                                                               
P(S_{1},S_{2})=\frac{1}{Z} T(S_{1},S_{2})T^{N-1}(S_{2},S_{1}).                 
\label{e18}                                                                    
\end{equation}
We can use the spectral development of the matrix $T$ given by
\begin{equation}                                                               
T(S_{1},S_{2})=\sum_{N}\phi_{k}(S_{1})\lambda_{k}\phi_{k}^{*}(S_{2}),          
\label{e19}                                                                    
\end{equation}
where $\phi_{k}(S_{1})$ is the normalized eigenvector of $T$ and
$\lambda_{k}$ is the corresponding eigenvalue, to write the expressions
\begin{equation}                                                               
Z =  \sum_{k} \lambda_{k}^{N} ,                                                
\label{e20}                                                                    
\end{equation}

\begin{equation}                                                               
P(S_{1})=\frac{1}{Z} \sum_{k}                                                  
\phi_{k}(S_{1})\lambda_{k}^{N}\phi_{k}^{*}(S_{1}),                              
\label{e21}                                                                    
\end{equation}
and
\begin{equation}                                                               
P(S_{1},S_{2})=\frac{1}{Z} T(S_{1},S_{2})                                      
\sum_{k} \phi_{k}(S_{2})\lambda_{k}^{N-1}\phi_{k}^{*}(S_{1}).                 
\label{e22}                                                                    
\end{equation}
In the limit $N \rightarrow \infty$, Eqs. (\ref{e20}), 
(\ref{e21}) and (\ref{e22}) become $\lambda_{0}^{N}$,
$P(S_{1})=\phi_{k}(S_{1})\phi_{k}^{*}(S_{1})$ and 
$P(S_{1},S_{2})=\frac{1}{\lambda_0} T(S_{1},S_{2})                              
\phi_{k}(S_{2}) \phi_{k}^{*}(S_{1})$, respectively. 
Putting $S_2 = S_1$ in the above expression, we arrive at
\begin{equation}                                                                
\lambda_{0}=\frac{ \langle T(S_{1},S_{1})\rangle}
{\langle\delta_{S_{1},S_{2}}\rangle},                     
\label{e17}                                                                    
\end{equation}
where  $\lambda_{0}$ is the largest eigenvalue 
of $T$ and $f$ becomes $f=-\frac{1}{\beta L} \ln \lambda_0$.
The quantity  $\delta_{S_{1},S_{2}}$  is the Kronecker delta for
$S_1$ and $S_2$ which is equal to $1$
when layers $S_{1}$ and $S_{2}$ are equal and zero otherwise. 
In the next section, we will write  down the transfer matrix 
$T(S_{1},S_{1})$ for the models studied here.
Since the  averages described above are evaluated over
all $N$ layers, from now on we are going to replace $T(S_{1},S_{1})$
by $T(S_{k},S_{k})$. 
More details about the transfer matrix method are found in Refs. 
\cite{sauerwein,fiore11,cluster2}.
\section{Models}
\subsection{Ising model}
The  Ising model \cite{ising} is defined as follows: 
each site of the lattice is attached by a spin variable
$\sigma$ that takes the values $\sigma_i = \pm 1$ according to whether the
spin is ``up'' or ``down'' (or equivalently, an 
 occupied or  empty site in a fluid jargon).  
The Hamiltonian is given by
\begin{equation}                                                              
{\cal H} = - J \sum_{<i,j>} \sigma_{i} \, \sigma_{j} - H \sum_i \sigma_i,      
\end{equation}
where $J$ is the interaction energy between
two nearest neighbor spins and $H$ is the magnetic field.
For $H=0$ and  low temperatures, the system
presents a phase coexistence between two
ferromagnetic phases which ends at the critical point
${\bar T}_c=2.269...$,where  ${\bar T}\equiv k_{B}T/J$.
The transfer matrix diagonal elements, which
will be used for obtaining the entropy in the CEM, reads
\begin{equation}                                                               
T(S_{k},S_{k}) = \exp\left[ \beta \, \big 
\{\sum_{i=1}^{L} J \, 
(1 + \sigma_{i,k} \, \sigma_{i+1,k}) + H \, \sigma_{i,k} \big \} \right],        
\end{equation}
where the sum is performed over a layer  with $L$ sites.

\subsection{BEG model}
The BEG model \cite{BEGMODEL} is a generalization 
of the Ising model, where
 each site is allowed to be empty or occupied by two distinct species. It 
is defined by the Hamiltonian
\begin{equation}                                                               
 {\cal H} = -\sum_{<i,j>} (J \sigma_{i} \, \sigma_{j} + K \sigma_{i}^{2} 
\sigma_{j}^{2}) +D \sum_{i}  \sigma_i^2,  
\label{e3}                                                                     
\end{equation}
where $\sigma_i=0$, if  the site $i$
is empty and $\pm 1$ if $i$ is occupied  
by one of the species. 
Parameters  $J$ and $K$ are interaction energies and $D$  denotes 
the chemical potential.
The BEG model displays a rather rich phase diagram, whose
features  depend on 
the ratio ${\bar K}\equiv K/J$. As far as the regime 
${\bar K}>-0.5$ and low temperatures is concerned, 
the system displays
 liquid ($\rho \neq 0$) and gas phases ($\rho=0$) for high and low
chemical potentials, respectively 
(or equivalently for low and high values of 
${\bar D}\equiv D/J$, respectively).
For ${\bar T}=0$, 
the liquid-gas phase coexistence takes place at 
 ${\bar D}^{*}=z({\bar K}+1)/2$, where $z$ is the coordination number. 
The transfer matrix, that
will be used for evaluating the entropy, 
is given by
\begin{eqnarray}                                                                
T( S_{k},S_{k}) = \exp\Big[ \beta \sum_{i=1}^{L}                                
\Big(  J \, \sigma_{i+1,k} \sigma_{i,k}                                 
\nonumber \\                                                                   
+ (J - D + K \, (1 + \sigma_{i+1,k}^{2})) \, \sigma_{i,k}^{2}                    
\Big) \Big].                                                                    
\end{eqnarray}

\subsection{Bell-Lavis model}
The  Bell-Lavis (BL) model is
defined on a triangular lattice where each site may be
empty ($\sigma_i=0$) or occupied by a water molecule ($\sigma_i=1$). 
Each  particle has two orientational states,
that may be described
in terms of bonding and inert ``arms'' 
$\tau_{i}^{ij}$, which   take the values
 $\tau_{i}^{ij} = 0$ or $\tau_{i}^{ij} = 1$ when
the arm is inert or bonding, respectively. 
Two nearest neighbor molecules interact via  van der Waals 
$\epsilon_{vdw}$
 and  hydrogen bond $\epsilon_{hb}$ 
energies whenever they are adjacent and 
point out their arms to each other ($\tau_{i}^{ij}\tau_{j}^{ji} = 1$),
respectively.
The BL model is defined by the following Hamiltonian  
\begin{equation}                                                               
{\mathcal H} = -\sum_{<i,j>} \sigma_{i} \, \sigma_{j} \,                  
(\epsilon_{hb} \, \tau_{i}^{ij} \, \tau_{j}^{ji} +                    
\epsilon_{vdw}) - \mu \sum_{i} \sigma_{i},                                     
\label{hambl}                                                                  
 \end{equation}
where  $\mu$ is the chemical potential. The BL model also displays a rich 
phase diagram, whose  
features  depend on the ratio $\zeta=\epsilon_{vdw}/\epsilon_{hb}$.  
In particular, for $\zeta=0.1$
one has three phases, denoted gas, low-density-liquid (LDL)
and  high-density-liquid (HDL)  \cite{bell,fiore-m}.
As in the BEG model, the gas phase is devoided by molecules, whereas
the LDL phase is characterized
by a honeycomb like structure, with density 
of particles $\rho$ and hydrogen bonds 
 per molecule $\rho_{hb}$  given by
$\rho=2/3$ and $\rho_{hb}=3/2$, respectively. 
In the HDL phase, the lattice
is filled by molecules, $\rho=1$, and the hydrogen 
bonds density per molecule
is also given by $\rho_{hb}=1$.
For  low values of ${\bar \mu}\equiv \mu/\epsilon_{hb}$, the system is 
constrained in the gas phase.
By increasing ${\bar \mu}$
for  fixed low  ${\bar T}$, a first  
 transition between the gas 
and the LDL phase occurs. 
By increasing further ${\bar \mu}$ a  second 
 transition, from the phase LDL to the 
HDL, takes place.
At ${\bar T}=0$, both transitions
are first-order and 
occurs at ${\bar \mu}^{*} = -3 \, (1+\zeta)/2$ and 
${\bar \mu}^{*} = -6 \, \zeta$, respectively.
For ${\bar T} \neq 0$ the former phase
transition  remains first-order, 
whereas the latter  becomes second-order \cite{fiore-m}. For $\zeta=0.1$, 
the second-order and first-order lines meet in a tricritical point.

The transfer matrix that
will be used for evaluating the entropy in the CEM is given by
\begin{eqnarray}                                                                
T(S_{k},S_{k})=
\exp\Big[\sum_{i=1}^{L} \{                                             
\sigma_{i,k} \, (\sigma_{i,k} + 2 \sigma_{i+1,k}) \\                              
(\epsilon_{vdw} + \epsilon_{hb} \, \tau_{i,k} \,                                 
\tau_{i+1,k} + \mu) \} \Big]\nonumber.
\end{eqnarray}

\section{Numerical results}

\subsection{Ising model}
\begin{figure}
\setlength{\unitlength}{1.0cm}
\includegraphics[scale=0.36]{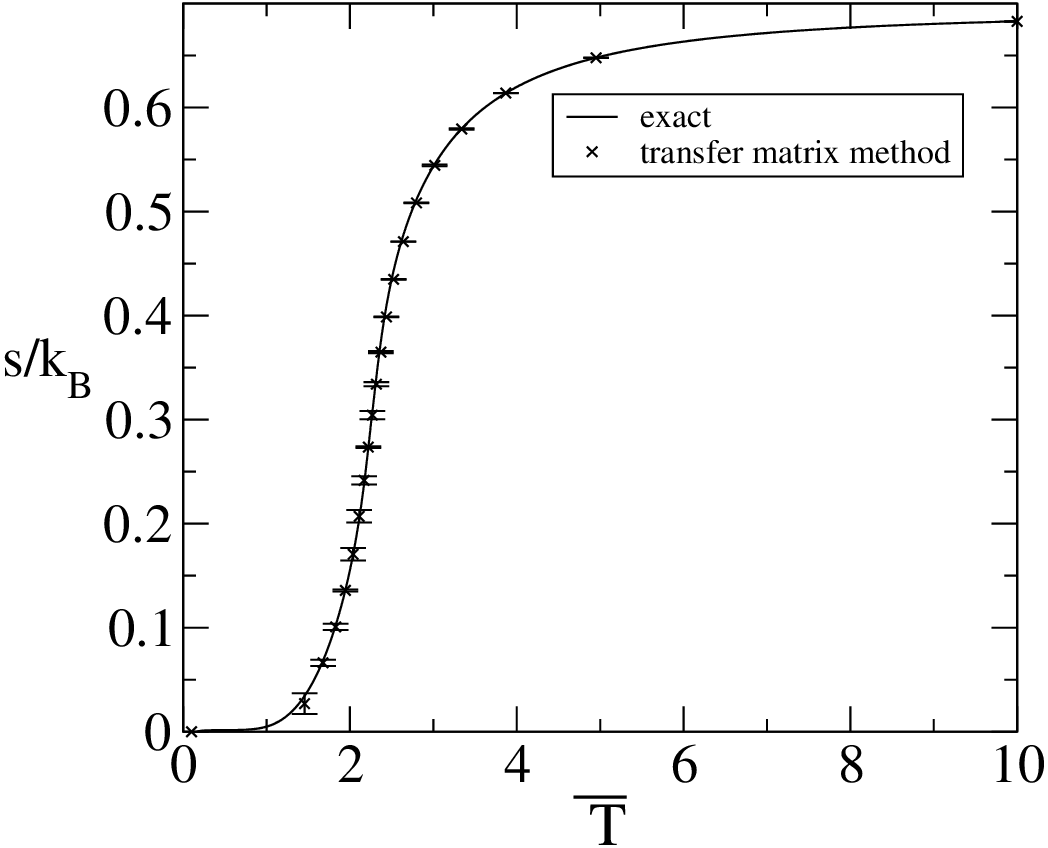}                  
\caption{Entropy per site $s/k_{B}$ versus temperature ${\overline T}$ for the 
Ising model. We have used $L=N=20$ and CE intervals. 
The result obtained via the transfer matrix method (stars) 
is compared with the exact result (solid line).}
\label{fig1}
\end{figure}

We have performed numerical simulations of the Ising model
in a square lattice of size $L=N=20$ using periodic boundary conditions.
We have discarded   $3 \times 10^{5}$ MC steps, in order to
 equilibrate the system and   $3 \times 10^{6}$ MC to evaluate the 
appropriate quantities.
In Fig. \ref{fig1}, we compare the entropy $s/k_B$ 
obtained via the transfer matrix method with 
 exact results \cite{ferdinand}.
We have considered a set of $R=21$ replicas 
ranged from ${\bar T}_1=0.1000$ to ${\bar T}_{21}=10.0000$ 
distributed according to the CEM.
The  agreement between results support the adequacy of the 
present procedure for obtaining $s$, 
which will be used for estimating the
temperature set. 

\begin{figure}
\setlength{\unitlength}{1.0cm}
\includegraphics[scale=0.4]{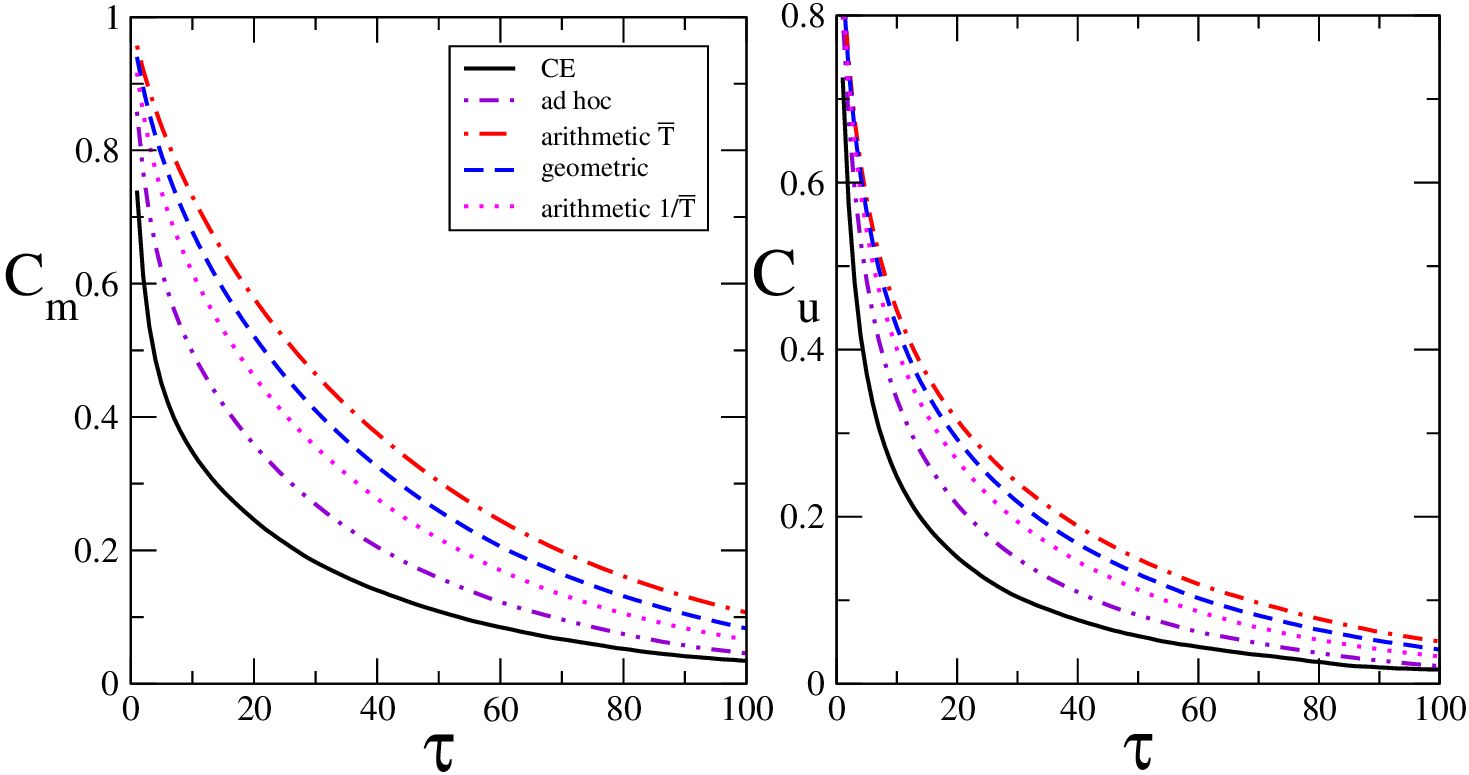}                                    
\caption{Time displaced auto-correlation functions $C_{m}$ 
and $C_{u}$ versus
time $\tau$ (in MC steps) for the Ising model at the critical
point for
$N=L=20$ and  different schemes of temperature selection.}
\label{fig2}
\end{figure}

In Table \ref{table1} we 
 compare our temperature estimates with those obtained by 
Sabo {\it et al}. \cite{sabo} by means of numerical
integration of the specific heat $C_v$. 
The data are shown to be in good overall agreement, even though 
some small discrepancies can be observed at certain temperatures. 
They may be explained by either numerical uncertainties in the 
transfer matrix method, or by uncertainties in the integrations of $C_v$, 
or by derivations of $u$ in the exact method, or even by all these 
sources together.
\begin{table}
\begin{ruledtabular}
\begin{tabular}{ccccc}
CEM (exact) & CEM-S (exact) & CEM (20) & CEM-S (20)  \\
\hline
0.1000 &0.1000& 0.1000 &0.1000    \\
1.4683 &1.4688&1.4551 &1.4635     \\
1.6867 &1.6866&1.6769 &1.6820    \\
1.8360 &1.8373&1.8289 &1.8332   \\
1.9524 &1.9538&1.9457 &1.9496    \\
2.0464 &2.0478&2.0379 &2.0433    \\
2.1236 &2.1250&2.1107 &2.1201    \\
2.1865 &2.1879&2.1696 &2.1836     \\
2.2363 &2.2374&2.2200 &2.2386    \\
2.2697 &2.2702&2.2681 &2.2883   \\
2.3048 &2.3064&2.3170 &2.3365    \\
2.3580 &2.3603&2.3711 &2.3896    \\
2.4288 &2.4321&2.4374 &2.4518    \\
2.5208 &2.5265&2.5242 &2.5341     \\
2.6408 &2.6473&2.6401 &2.6464    \\
2.8007 &2.8115&2.7977 &2.8019   \\
3.0219 &3.0372&3.0170 &3.0220    \\
3.3481 &3.3708&3.3393 &3.3483    \\
3.8852 &3.9414&3.8687 &3.8847    \\
4.9930 &5.1247&4.9505 &4.9906    \\
10.000 & 10.000&10.000 &10.000  
\end{tabular}
\end{ruledtabular}
\caption{\label{table1} Temperature CEM set for the Ising
model obtained from 
the present approach (first and third columns) and
by Sabo et. al \cite{sabo} (second and fourth columns).}
\end{table}
In the first comparison among temperature schedules, 
 we investigate the decay of time-correlation displaced 
functions $C_{q}$ at the critical point. 
This study is motivated by the fact  that 
numerical simulations of second-order phase transitions
via conventional algorithms are affected by 
a slow decay of $C_{q}$
(critical slowing down). On the other hand,
  cluster algorithms \cite{sw} 
reduce drastically this effect. This  suggests that  the 
analysis of $C_{q}$ may be a good measure
for the comparison of different criteria used in the PT.
The auto-correlation function 
$C_{q}$ of a given quantity $q$ at the time $\tau$ is given by
\begin{equation}                                                         
C_{q}(\tau) =                                                            
\langle (q(t) - {\bar q})(q(t + \tau) - {\bar q}) \rangle/\sigma_q^2,    
\label{acf}                                                              
\end{equation}  
where $\bar q$ is the mean value and $\sigma_q$  the
 variance, respectively.
In Fig. \ref{fig2} we plot  $C_{q}$ for
the thermodynamic quantities $m=\langle \sigma_{i}\rangle$ 
(magnetization per site) and $u=\langle {\cal H}\rangle$ (total
energy per site) for all distributions described above.  
We have considered  $R=6$ replicas
ranged from  ${\bar T}_1=2.269$ to ${\bar T}_4=3.660$,
whose temperature set is showed in Table \ref{table2}
for the CEM and {\it ad hoc} distributions.
By putting the temperature set  into an array, 
we have considered exchanges between every adjacent 
and non-adjacent  (here between every second and third) temperatures.
\begin{table}
\begin{tabular}{cc}\hline
CEM & {\it ad hoc}   \\ \hline
2.269 & 2.269  \\
2.345 & 2.400   \\
2.464 & 2.580  \\
2.658 & 2.830  \\
2.996 & 3.170  \\
3.660 & 3.660 \\\hline
\end{tabular}
\caption{\label{table2} Temperature set of $R=6$ replicas
for the Ising model obtained
for the CEM and {\it ad hoc} distributions.}
\end{table}

Although  all schemes give equivalent
results for the steady state (for $u$ and $m$), the quantity 
$C_{q}$  decays faster  within the CEM. 
As it will be shown later, a similar behavior
is verified when one calculates $C_{q}$   
at the critical point for the BL model. In particular,
by allowing only adjacent exchanges, both
$C_{m}$  and  $C_{u}$ also decay faster  with the CEM than other schedules.
\subsection{BEG model}
\begin{figure}
\setlength{\unitlength}{1.0cm}
\includegraphics[scale=0.4]{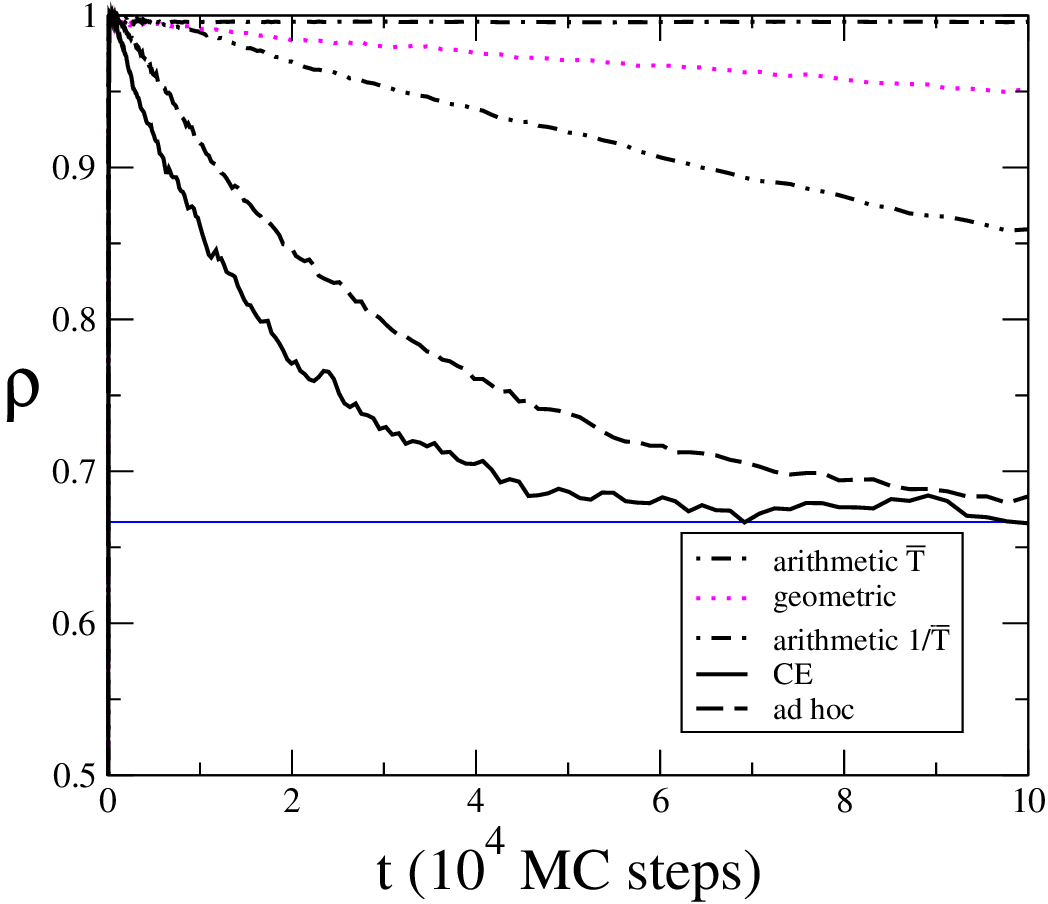}                
\caption{Time decay of the order parameter $\rho$ versus
time  (in MC steps) for the BEG model
at the phase coexistence point (${\bar D}^{*},{\bar T_1}$)=($8.0000,0.5000$)
for  $N=L=20$ and 
different distribution of temperatures. The tie line for $\rho=2/3$
denotes its stationary value $\rho_0$.}
\label{fig3}
\end{figure}

\begin{figure}
\setlength{\unitlength}{1.0cm}
\includegraphics[scale=0.4]{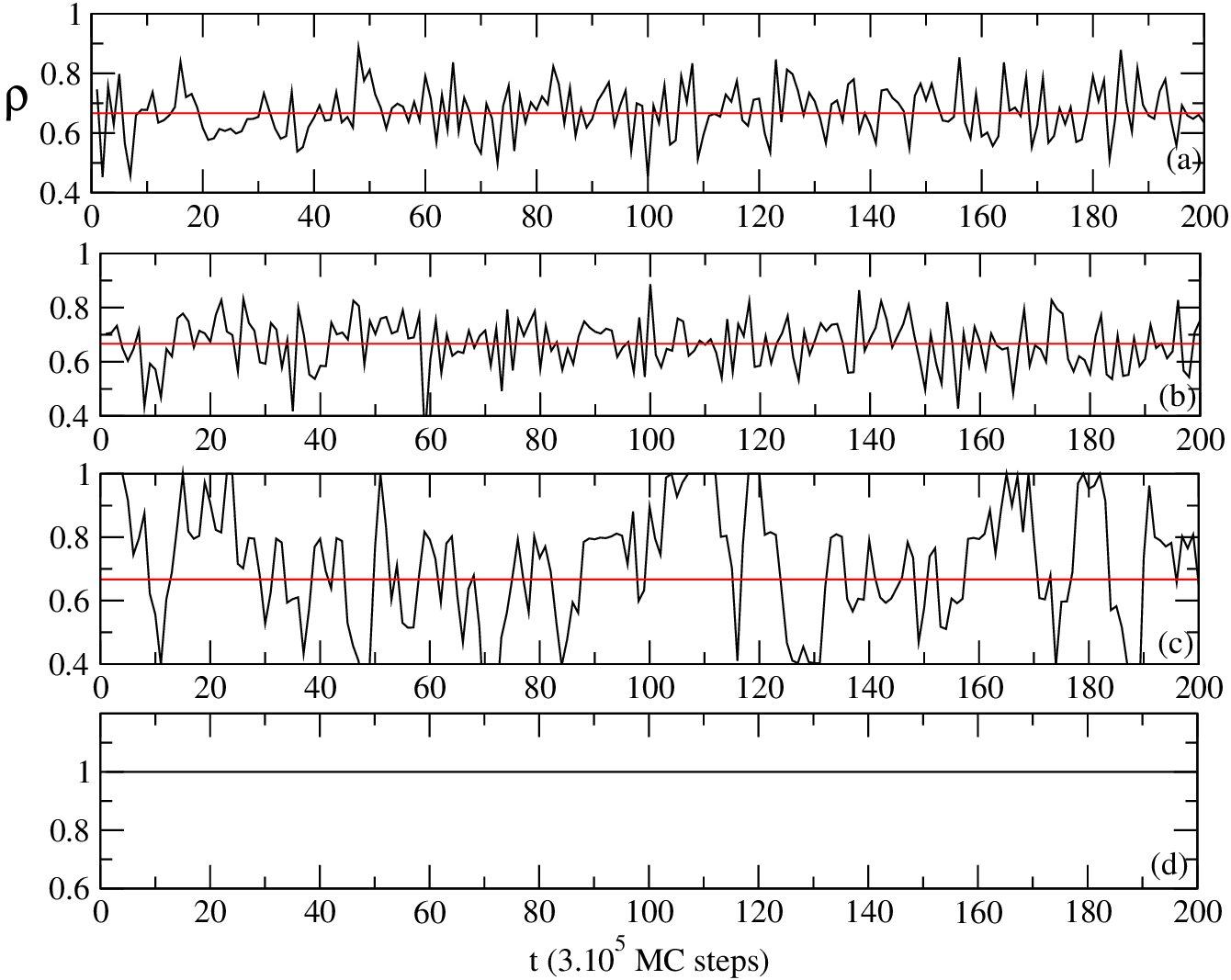}
\caption{The density $\rho$ versus the time $t$ 
at the phase coexistence  
(${\bar D}^{*},{\bar T_1}$)=($8.0000,0.5000$)  
 after discarding 
$10^{6}$ initial MC steps for the CEM, {\it ad hoc}, geometric
and arithmetic $1/{\bar T}$ distributions  (graphs $(a),(b),(c)$ and $(d)$, 
respectively). The tie lines in graphs $(a),(b),(c)$ denotes                       
the stationary $\rho_0$.}
\label{fig4-5}
\end{figure}

We have performed numerical simulations for the BEG
model in a square lattice of size $L=N=20$ using periodic boundary conditions.
We  focus on the analysis for  ${\bar K}=3$, ${\bar H}=0$ and 
the low temperature ${\bar T}_1=0.5000$. 
In this case, a  first-order
phase transition between the liquids
 and the gas phase  takes place at  ${\bar D}^{*}=8.0000(1)$ 
\cite{fiore8}.
Since the probability distribution at  discontinuous 
transitions exhibit two peaks (corresponding
to each phase), conventional Monte Carlo algorithms  are not efficient 
at low temperatures, since 
the system requires a long time to pass from one peak to the other.
In extreme cases, the peaks
are separated by very high barriers and 
the system may get trapped in a given phase 
along the whole simulation and in this case it will be not ergodic.
With these concepts in mind, we consider two  analyses 
at the phase coexistence: the time evolution
of the order parameter $\rho=\langle \sigma_{i}^{2} \rangle$ 
toward its equilibrium value $\rho_0=2/3$ \cite{ref20} starting from a 
non typical configuration   and the tunneling
between the phases at the coexistence after 
discarding  sufficient MC steps. The latter study will be 
carried out by measuring the fluctuation of $\rho$ around 
 $\rho_0$, since  the trapping of the system
in a given phase or in a mestastable state is expected to be 
signed by  no relevant change of $\rho$.
We have distributed  
 temperatures between
${\bar T}_1=0.5000$ and ${\bar T}_{R}=2.0000$,
whose entropies  per site are given by $s_1/k_{B}=5\times 10^{-6}$ and
$s_{R}/k_{B}=0.4971(1)$, respectively.  By
using in all cases a set of $R=6$ replicas,  
the CEM criterion leads to the intermediate
temperatures  ${\bar T}_2=1.5550$, 
${\bar T}_3=1.7650$, ${\bar T}_4=1.8780$ and 
${\bar T}_5=1.9400$. As for the Ising model, 
we have considered exchanges between every adjacent
and non-adjacent  (here between every second and third) temperatures.
In Fig. \ref{fig3}, we plot
the time evolution of  $\rho$ starting from a lattice
filled with particles.
Note  that by choosing the temperatures according to the CEM,
the convergence of $\rho$ toward 
 $\rho_{0}$  is faster  than for other schemes.
Although  the results obtained
for the {\it ad hoc} and  CE cases are close,
only in the latter scheme
the system   reached the steady state until $10^{5}$ MC steps. 
In Fig. \ref{fig4-5} we plot the quantity $\rho$ versus the time $t$ 
(in MC steps)  for all procedures after discarding  $1\times 
10^{6}$ initial MC steps.
We considered an initial configuration
filled with particles and the densities are evaluated 
each $3 \times 10^{5}$ MC steps.
With the exception of the arithmetic $1/{\bar T}$ criterion, 
where the simulation gets trapped in the liquid phase 
($\rho \approx 1$) the whole time of simulation, 
the system is able to cross the  
free-energy barriers properly in the other cases. 
This   can be viewed by the fluctuations around 
its equilibrium value $\rho_0$, whose  density averages
${\bar \rho}$ are consistent with $\rho_0=2/3$ 
for arithmetic ${\bar T}$, geometric and CEM. 
In the next application, we shall see that 
the choice of the temperature interval
will have more influence on the tunneling.

\subsection{Bell-Lavis model}
\begin{figure}
\setlength{\unitlength}{1.0cm}
\includegraphics[scale=0.4]{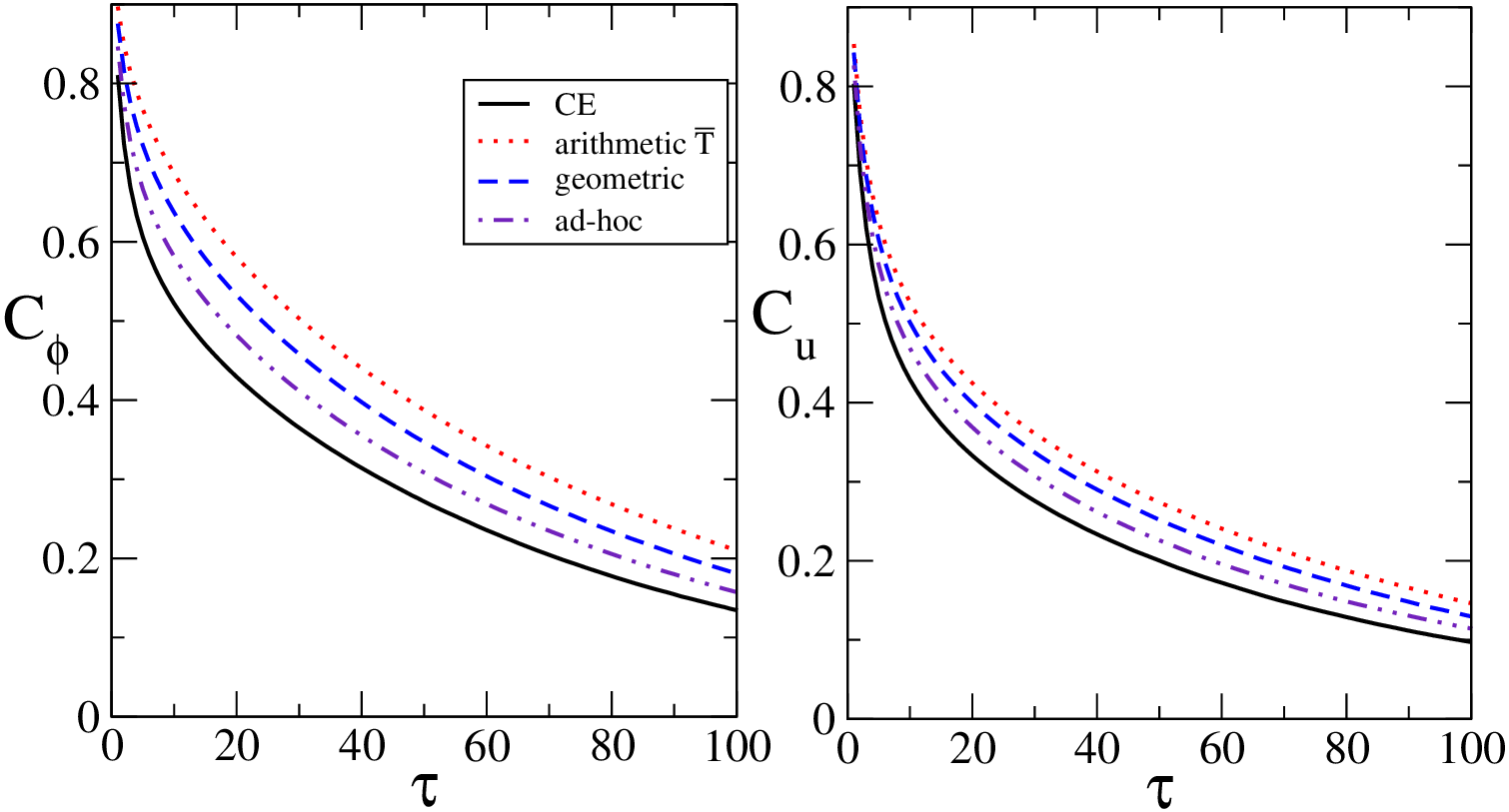}                                   
\caption{Time displaced auto-correlation functions $C_{\phi}$ 
and $C_{u}$  versus
time $\tau$ (in MC steps) for the BL model at the critical 
point  $({\bar \mu_c},{\bar T_c})=(-1.000,0.430)$  for
$N=L=18$ and   different temperature schedules. Results
for the arithmetic $1/{\overline T}$ 
have been omitted as they are similar
to the {\it ad hoc} ones.}
\label{fig4c}
\end{figure}

In the last part of this paper, we study the BL model
in triangular lattice of size $L=N=18$ using periodic boundary conditions. 
First, we repeat the analysis performed for the Ising model 
in the  LDL-HDL second-order transition. 
We recall that the   density of 
particles $\rho$ is not the order-parameter $\phi$, since $\rho \neq 0$ 
in both liquid phases. A previous study
\cite{fiore-m} showed that  the
appropriate $\phi$  is the difference 
 between the fullest  $\rho_i$  and the emptiest $\rho_j$  
density sublattices given by $\phi=\rho_i - \rho_j$. 
In Fig. \ref{fig4c} we plot the auto-correlation functions 
$C_{\phi}$ and $C_{u}$  for all distributions. 
In particular, for the critical point located at 
$({\bar \mu_c},{\bar T_c}$)=($-1.000,0.430$),
we distributed $R-2=4$ replicas between ${\bar T}_1=\bar T_c$ and 
${\bar T}_{R}=0.730$, 
whose entropies per site are $s_1/k_B=0.5550(2)$ and 
$s_R/k_B = 0.8763(2)$, respectively. 
We have also considered  exchanges between every adjacent
and non-adjacent  (here between every second and third) temperatures.
Table \ref{table3}
 shows the temperature set  for the CEM and
{\it ad hoc} cases.
\begin{table}
\begin{tabular}{cc}\hline
CEM & {\it ad hoc}   \\ \hline
0.4300 & 0.430  \\
0.4601 & 0.468   \\
0.5022 & 0.518  \\
0.5589 & 0.576  \\
0.6340 & 0.645  \\
0.7300 & 0.730 \\\hline
\end{tabular}
\caption{\label{table3} Temperature set   
of $R=6$ replicas                                         
for the BL model at  ${\bar \mu_c}=-1.000$ obtained                       
for the CEM and {\it ad hoc} distributions.}
\end{table}
As in the Ising model, 
$C_{\phi}$ and $C_{u}$  also decay faster
at the critical point when
temperatures are chosen using the CEM. 
 Repetition for other critical points 
 leads to the same conclusion. 

\begin{figure}
\setlength{\unitlength}{1.0cm}
\includegraphics[scale=0.4]{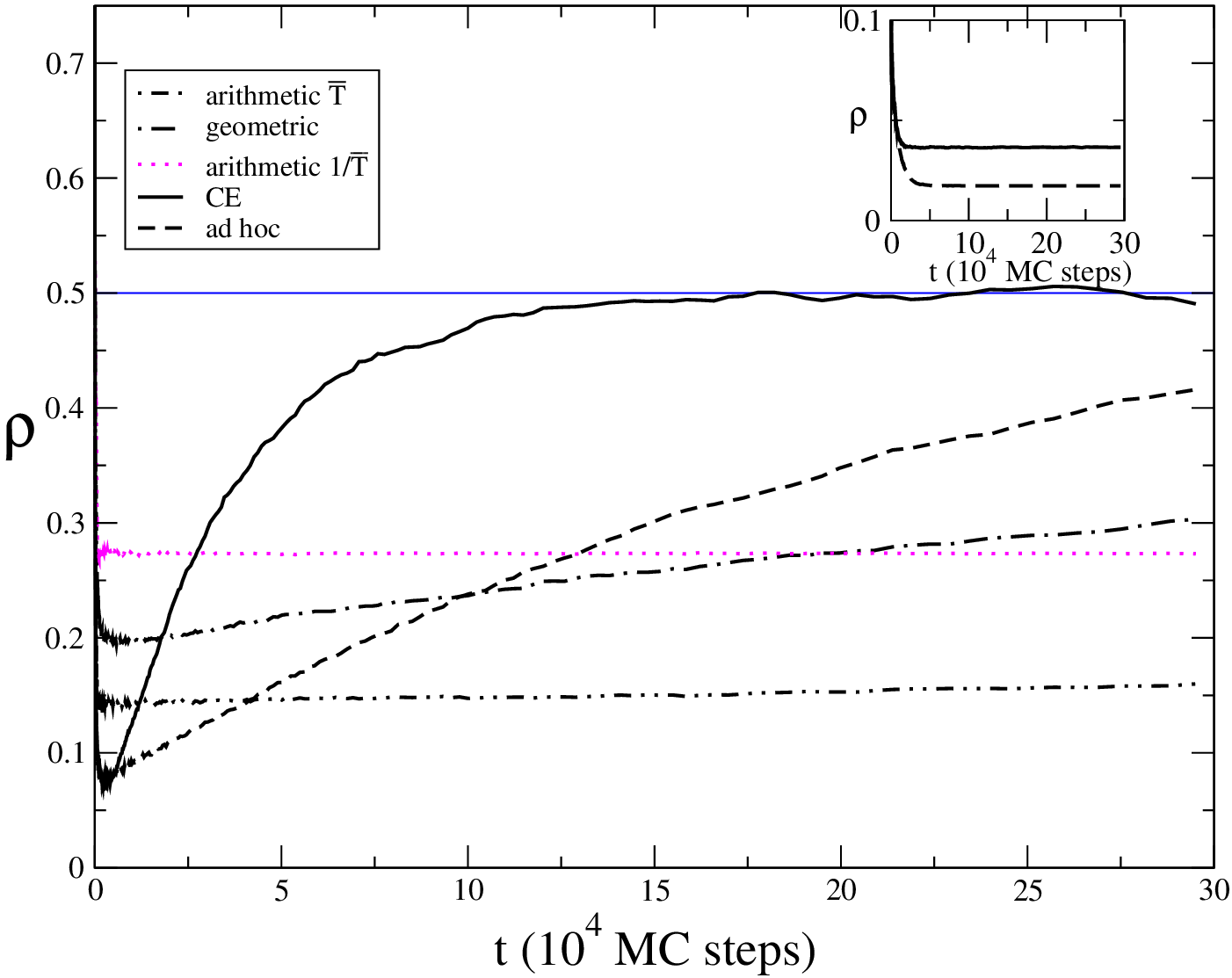} 
\caption{Time decay of the order parameter $\rho$ versus
time  (in MC steps) for the BL model  
at the phase coexistence (${\bar \mu}^{*},{\bar T_1}$)=($-1.6500,0.1000$)
for $N=L=18$ and different distribution of temperatures. 
The tie line for $\rho=1/2$  denotes its stationary value $\rho_0$.
In the inset we plot, for the {\it ad hoc} and CEM schedules, 
the decay of $\rho$ considering only adjacent swaps.}
\label{fig5}
\end{figure}

In addition to the previous study, 
we  also investigate  the first-order phase transition 
gas-LDL occurring at low temperatures.
Numerical simulations 
have been carried out at ${\bar T_1}=0.1000$. For
this temperature, the phase transition takes place
at ${\bar \mu}^{*}=-1.6500(1)$, 
which is identical (up to the fourth decimal
level) to the transition point  ${\bar \mu}=-1.65$ 
calculated  at ${\bar T}=0$.
In  Fig. \ref{fig5},
we plot the time evolution of the density of molecules
$\rho=\langle \sigma_{i} \rangle$ starting from an initial
configuration filled by molecules. We consider extreme temperatures 
${\bar T}_1=0.1000$ and ${\bar T}_{R}=0.4200$, with corresponding
entropies per site  given by 
$s_1/k_{B}=10^{-5}$ and
$s_{R}/k_{B}=0.4604(1)$, respectively. By considering 
in all cases a set of $R=6$ replicas we have, for the CEM case, the 
intermediate temperatures   ${\bar T}_2=0.2837$, 
${\bar T}_3=0.3456$, ${\bar T}_4=0.3756$ and
${\bar T}_5=0.3973$.
As for the BEG model,
with the CEM the system crosses the entropic
barriers more frequently than with other criteria, 
which can be identified by the faster convergence of $\rho$ 
toward its equilibrium value $\rho_{0} \approx 1/2$ 
\cite{fiore11,ref20}. On the other hand, for
the other procedures, the system remains a larger number 
of MC steps trapped in metastable configurations and 
until $3 \times 10^{5}$ MC steps the density  
has not yet converged  to $\rho_{0}$. When we consider only adjacent replica
exchanges, the system gets trapped in metastable states for
all distributions. In the inset of Fig. \ref{fig5} we plot the decay
of $\rho$ only for the CEM and {\it ad hoc} schedules.

\begin{figure}
\setlength{\unitlength}{1.0cm}
\includegraphics[scale=0.4]{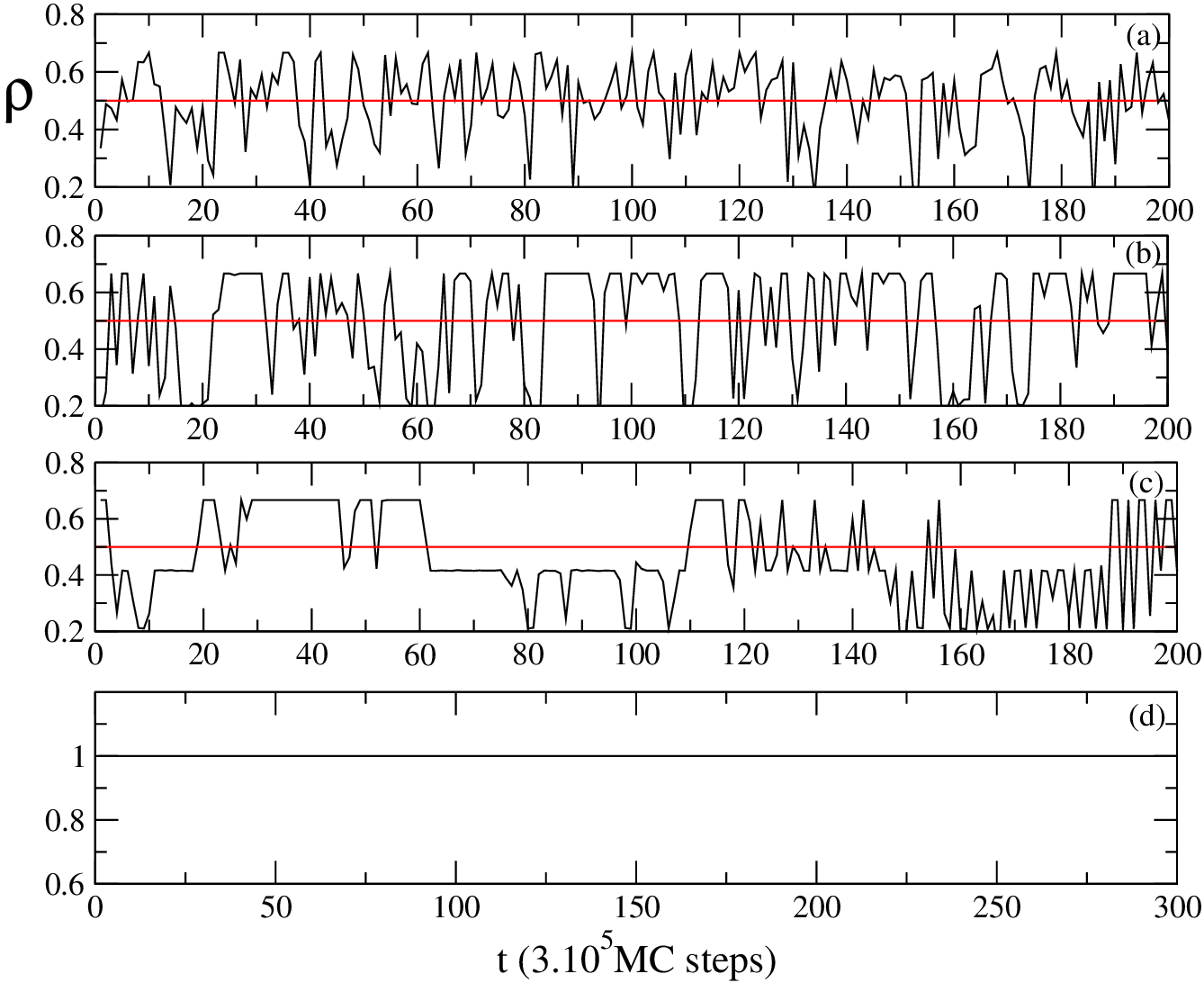}
\caption{The density $\rho$ versus the time $t$ 
at the phase coexistence  (${\bar \mu}^{*},{\bar T_1}$)=($-1.6500,0.1000$) 
after discarding               
$10^{6}$ initial MC steps for the CEM, {\it ad hoc}, geometric   
and arithmetic $1/{\bar T}$ distributions 
(graphs $(a),(b),(c)$ and $(d)$,         
respectively) and $N=L=18$.  The tie lines in graphs $(a),(b),(c)$ denotes
the stationary $\rho_0$.}
\label{fig6-5}
\end{figure}

We also  show in Fig.
\ref{fig6-5} the density $\rho$
versus $t$ also starting from an initial configuration
filled by  molecules  after discarding $10^{6}$ initial MC steps. 
In the BL model, only with the CEM and {\it ad hoc} criteria the system
crosses frequently the  free-energy barriers, 
although  the tunneling is rather more
frequent with the CEM than the {\it ad hoc}  
distribution. Only in these cases the
tunneling between the different phases give
an  equilibrium value ${\bar \rho}=0.504(3)$,
consistent with $\rho_0=1/2$.
On the other hand,
for arithmetic ${\bar T}$ (not shown) and geometric schedules the tunneling is 
much less frequent than the CEM ones, 
whereas  for the  arithmetic ${1/\bar T}$ the 
systems  gets trapped the whole simulation in a metastable state generated
by the initial configuration. 

As a final remark, it is worth mentioning that the 
observed differences in the results yielded by the 
various temperature schemes are less pronounced
in discontinuous transitions occurring
at high temperatures. In addition,  
by increasing both the number of replicas 
and the degree of non-adjacent exchanges
 the results also become closer.

\section{Discussion and conclusion}
We  presented a comparative study among different protocols 
for the choice of
the temperature set in the parallel tempering method.
We focused our attention on five
criteria denoted arithmetic, geometric progressions 
in the temperature, arithmetic progression in the inverse temperature,
{\it ad hoc} distribution and the constant entropy method (CEM). 
In this last case,
we considered an alternative direct MC method
 for evaluating the entropy
which avoids  numerical
integrations of the specific heat or other thermodynamic quantities.
We have considered 
rather few number of replicas ($R=6$
for continuous and discontinuous phase transitions) 
and adjacent and non-adjacent replica exchanges. 
Different   systems  undergoing first and second-order
phase transitions have been undertaken. In all cases, 
the temperature selection via the difference
of entropy  method revealed more advantageous. More specifically,
at the criticality (where configurations generated
by standard  algorithms become strongly correlated)  
the time displaced correlation functions decay faster when temperatures
are  chosen with the CEM.
This behavior can be  understood 
that near criticality (where a small 
change of temperature  provoke a large change
of entropy) the CEM gives more concentrated
intermediate temperatures than all distributions. Thus, replicas at
the lowest temperature (${\bar T}={\bar T_1}$) 
display a  larger  probability of exchanging configurations
than the other  cases.
Since   the time correlation decays faster for ${\bar T}>{\bar T_c}$ than
${\bar T}={\bar T_c}$, the more frequent exchanges provide the system at
${\bar T_1}$ decays faster.
For discontinuous  transitions 
at low temperatures, where high entropic barriers do not allow 
the system to cross  the phase frontiers properly (also when  simulated
by conventional algorithms)  and hence the choice
of the adjacent temperatures may play a crucial role,
the CEM has  also offered
a rather efficient recipe for determining the temperature set.
Within the CEM the lower temperatures are more sparse than with other schemes
and,  though unlikely, a successful replica exchange allows
the system  to evolve to configurations which are able to cross the high
free energy barriers faster than  other distributions.
We have also distributed temperatures following an {\it ad hoc} scheme, 
in such a way that the exchange probability between 
adjacent replicas was about 30$\%$. Although
this method has shown to be more efficient than arithmetic and geometric schedules at the phase
transition, it is inferior than the CEM.
In summary, our comparative study ellects the CEM as 
an useful tool for obtaining the temperature
schedule to be used in  numerical simulations of phase transitions
through the parallel tempering method.
\section{Acknowledgments} 
I acknowledge  Renato M. \^Angelo, Mauricio Girardi,
Sergio D'Almeida Sanchez and Marcos G. E. da Luz for  critical
readings of this manuscript and the financial support from CNPQ.
  


\begin{thebibliography}{99}   

\bibitem{nemoto} K. Hukushima and K. Nemoto, J. Phys. Soc. Jpn. 
{\bf 65}, 1604 (1996); C. J. Geyer, 
{\it Markov-Chain Monte Carlo maximum Likehood}, 
Comp. Sci. and Stat.,  p. 156 (1991).

\bibitem{parisi} E. Marinari and G. Parisi, Europhys. Lett. {\bf
    19}(6), 451 (1992).

\bibitem{spinglass} K. Binder and W. Kob, {\it Glassy Materials and
    Disordered Solids: An Introduction to their Statistical Mechanics}
  (World Scientific, Singapoure, 2005).

\bibitem{proteins}  J. Skolnick and A. Kolinski,
  Comput. Sci. Eng. {\bf 3}(9/10), 40 (2001).

\bibitem{kofke} D. A. Kofke, J. Chem. Phys {\bf 117},
6911 (2002).


\bibitem{predescu} C. Predescu, M. Predescu and C. Ciobanu,
  J. Chem. Phys. {\bf 120}, 4119 (2004);
  J.  Phys. Chem, B {\bf 109}, 4189 (2005).

\bibitem{kone} A. Kone and D. A. Kofke, J. Chem. Phys {\bf 122},
206101 (2005).

\bibitem{helmut} H. G. Katzgraber, S. Trebst, D. A. Huse and
 M. Troyer, J. Stat. Mech. {\bf 3},  P031018  (2006).

\bibitem{sabo} D. Sabo, M. Meuwly, D. L. Freeman and J. D. Doll,
J. Chem. Phys {\bf 128}, 174109 (2008).
\bibitem{sabo2} D. Sabo, private communication (2011).


\bibitem{calvo} F. Calvo, J. Chem. Phys.
{\bf 123}, 124106 (2005).

\bibitem{sauerwein}
R. A. Sauerwein and M. J. de Oliveira,
Phys. Rev. B, {\bf 52}, 3060 (1995).

\bibitem{ising} E. Ising, Z. Phys. {\bf 31}, 253 (1925).

\bibitem{BEGMODEL}   M. Blume, V. J. Emery, and R. B. Griffiths,
  Phys. Rev. A {\bf 4}, 1071 (1971),  W. Hoston and
  A. N. Berker, Phys. Rev. Lett. {\bf 67}, 1027 (1991).

\bibitem{bell}
G. M. Bell and D. A. Lavis, J. Phys. A {\bf 3}, 568 (1970).

\bibitem{metr} N. Metropolis, A. W. Rosenbluth, M. N.  Rosenbluth and
  A. H. Teller, J. Chem. Phys. {\bf 21}, 1087 (1953).


\bibitem{juan} J. P. Neirotti, F. Calvo, D. L. Freeman and J. D. Doll, 
J. Chem. Phys. {\bf 112}, 10340 (2000).

\bibitem{juan1} F. Calvo, J. P. Neirotti, D. L. Freeman and
  J. D. Doll, J. Chem. Phys. {\bf 112}, 10350 (2000).



\bibitem{fiore8}
C. E. Fiore, Phys. Rev. E  {\bf 78}, 041109 (2008).

\bibitem{fiore10}
C. E. Fiore and M. G. E. da Luz, Phys. Rev. E  {\bf 82}, 031104 (2010).

\bibitem{fiore11}  
C. E. Fiore and M. G. E. da Luz, J. Chem. Phys {\bf 133}, 104904 (2010). 
\bibitem{ferdinand}
B. Kaufman, 
Phys. Rev. {\bf 76}, 1232 (1949);
A. E. Ferdinand and M. E. Fisher, 
Phys. Rev. {\bf 185}, 832 (1969).
\bibitem{binder} K. Binder and D. W. Heermann, Monte Carlo Simulation in
Statistical Physics (Springer-Verlag, New York Berlin Heidelberg, 1992).
           
\bibitem{fiore-m} 
C. E. Fiore, M. M. Szortyka, M. C. Barbosa and V. B. Henriques,
J. Chem. Phys {\bf 131}, 164506 (2009).


\bibitem {sw}R. H. Swendsen and J. S. Wang, Phys. Rev. Lett. {\bf 58},
  86 (1987), U. Wolff, Phys. Rev. Lett {\bf 62}, 361 (1989).

\bibitem{fiore4}
C. E. Fiore, V. B. Henriques and M. J. de Oliveira,
J. Chem. Phys. {\bf 125}, 164509 (2006).

\bibitem{fernades-levin}
H. C. M. Fernandes, J. J. Arenzon and Y. Levin,
J. Chem. Phys. {\bf 126}, 114508 (2007).

\bibitem{cluster2}
C. E. Fiore and C. E. I. Carneiro,
Phys. Rev. E {\bf 76}, 021118 (2007).

\bibitem{ref20} For the BEG model, 
the equilibrium value for $\rho$ at the phase coexistence
 can be understood recalling that
 two liquid phases ($\rho \approx 1$) coexist
with one gas phase ($\rho \approx 0$).
Since their weights are equal (1/3), 
we have $\rho_{0} \approx 2/3$ for
any system size. A similar reasoning shows $\rho_{0} \approx 1/2$
for the BL model at the phase coexistence.





\end{thebibliography}
\end{document}